\documentclass[reprint,amsmath,amssymb,aps,prb,superscriptaddress]{revtex4-2}
\usepackage{graphicx}
\usepackage{dcolumn}
\usepackage{bm}
\usepackage[colorlinks=true, letterpaper=true, pdfstartview=FitV, linkcolor=blue, citecolor=blue, urlcolor=blue]{hyperref}
\usepackage{multirow}
\usepackage{xcolor}
\usepackage{amsmath}
\usepackage{amsfonts}
\usepackage{amssymb}
\usepackage{graphicx}
\usepackage[T1]{fontenc}
\usepackage{mathptmx}
\usepackage{booktabs}

\begin{document}

\title{Two-dimensional rare-earth Janus 2$\textit{H}$-Gd$\textit{XY}$ ($\textit{X}$,$\textit{Y}$=Cl, Br, I, $\textit{X}$$\neq$$\textit{Y}$) monolayers: Bipolar ferro-magnetic semiconductors with high Curie temperature and large valley polarization}

\author{Cunquan Li}
\affiliation{Key Laboratory of Display Materials and Photoelectric Devices, Ministry of Education, Tianjin Key Laboratory for Photoelectric Materials and Devices, National Demonstration Center for Experimental Function Materials Education, School of Material Science and Engineering, Tianjin University of Technology, Tianjin, 300384, China}

\author{Yukai An}
\email{ykan@tjut.edu.cn}
\affiliation{Key Laboratory of Display Materials and Photoelectric Devices, Ministry of Education, Tianjin Key Laboratory for Photoelectric Materials and Devices, National Demonstration Center for Experimental Function Materials Education, School of Material Science and Engineering, Tianjin University of Technology, Tianjin, 300384, China}

\date[]{}
\begin{abstract}
  Two-dimensional (2D) ferromagnetic semiconductors show great interest due to their potential applications for the nanoscale electronic devices. In this work, the Janus 2$H$-Gd$XY$ ($X$, $Y$=Cl, Br, I, $X$$\neq$$Y$) monolayers with rare-earth element Gd (4$f^{7}$+5$d^{1}$) are predicted by the first-principles calculations. Small exfoliation energy of less than 0.25 J/m$^{2}$ and excellent dynamical/thermal stabilities can be confirmed for the Janus 2$H$-Gd$XY$ monolayers, which exhibit the bipolar magnetic semiconductor character with high Curie temperatures above 260 K and large spin-orbit coupling effect, and can be further transformed into the half-semiconductor phase under proper tensile strains (5-6\%). In addition, the in-plane magnetic anisotropy can be observed in the 2$H$-GdICl and 2$H$-GdIBr monolayers. On the contrary, the 2$H$-GdBrCl monolayer exhibits perpendicular magnetic anisotropy character, which originates from the competition between Gd-$p$/$d$ and halogen atom-$p$ orbitals. Calculated valley optical actions of the Janus 2$H$-Gd$XY$ monolayers exhibit distinguished valley-selective circular dichroisms, which is expected to realize the special valley excitation by polarized light. Spontaneously valley-Zeeman effect in the valance band for the Janus 2$H$-Gd$XY$ monolayers induces a giant valley splitting of 60-120 meV, which is also robust against various external biaxial strains. Tunable valley degree of freedom in the Janus 2$H$-Gd$XY$ systems is very necessary for encoding and processing information.
\end{abstract}

\maketitle


\section{Introduction}
\indent Two-dimensional (2D) materials show an increasing interest owing to their broad application prospects in nanoscale electronic devices. It has been proved by the Mermin-Wagner theorem that 2D magnetic ordering can be prohibited due to the thermal fluctuations \cite{Mermin1966}. As a breakthrough development, intrinsic ferromagnetic (FM) ordering was experimentally discovered in the CrI$_{3}$ \cite{Huang2017} and VI$_{3}$ \cite{Kong2019} monolayers, which exhibit the Curie temperatures (T$_{c}$) of 45 K and 49 K, respectively. These studies attract much attention and further stimulate in looking for 2D magnetic materials at the nanoscale. Until now, significant progress has been made with 2D materials with magnetic ordering, while the rare intrinsic semi-conducting ferromagnetism and low T$_{c}$ limit their potential applications in spintronic devices \cite{Dayen2020, Chen2019, Das2004}. The T$_{c}$ of 2D materials can be effectively enhanced by applying the charge carrier doping \cite{Jiang2018}, biaxial strain \cite{Lv2019}, and external electric fields \cite{Li2019}, etc. However, these methods are still not the ideal solutions for the practical applications in spintronics. 

\indent Recently, transition metal dichalcogenides chlorides (TMDs) monolayers with hexagonal structure own a pair of non-equivalent energy valleys at the Dirac points of the Brillouin zone, which can be excited using left- or right-handed light and attract a great deal of attention. The valley, which indicates the maximum of valence band or minimum of conduction band, is a new degree of freedom of carriers besides the charge and spin. The two inequivalent valleys constitute a binary index for low energy carriers and this degree of freedom can be encoded and manipulated as an information carrier in valleytronics \cite{Mak2018}. Inspired with the analysis above, searching for intrinsic ferro-valley (ferromagnetic materials with spontaneous valley polarization) \cite{Tong2016} materials with high T$_{c}$ and large valley polarization are necessary for the candidates of valleytronics. In order to realize this, it is vital to choose a system with special electrons and orbitals. Recently, rare-earth element-based systems, especially for the Gd element, have become a hot topic of research. In fact, as early as 1965, Mee $et~al$ \cite{Mee1965} firstly synthesized the block GdI$_{2}$ with van der Waals structure. Very recently, GdI$_{2}$ monolayer with high T$_{c}$ of 241 K and sizable MAE was firstly predicted by Wang $et~al$ \cite{Wang2020}, and its large valley polarization (149 meV) and robustness to external biaxial strain were investigated by Feng $et~al$ \cite{Cheng2021}. Immediately after, the GdX$_{2}$ (X=I, F, Cl, Br) monolayers \cite{Sheng2022, Ding2021, Liu2021} were further predicted, especially that the GdCl$_{2}$ monolayer shows high T$_{c}$ of about 224 K and large perpendicular magnetic anisotropy (PMA) character. Experimentally, Janus (special types of materials which have two faces at the nanoscale) TMDs monolayers has been synthesized successfully in special conditions \cite{Lu2017, Zhang2017}. And, based on DFT calculations, some magnetic Janus monolayers, such as FeClBr \cite{Haili2021}, MnSSe \cite{Gang2022}, Cr$_{2}$X$_{3}$S$_{3}$ (X=Br, I) monolayers \cite{Wu2021}, etc., have been predicted. Due to the super-/direct-exchange interactions and spin-orbit coupling (SOC) in these 2D magnetic Janus systems, an intrinsic valley polarization can be observed in both valance and conduction bands. Furthermore, for the Janus 2$H$-GdClF monolayer with ferromagnetic ordering, it is noted that the tunable valley polarization can be observed and driven by external biaxial strains \cite{Guo2022}. In light of these key factors about the systems with Gd element mentioned above, the Janus 2$H$-Gd$XY$ monolayers can be expected as a ferrovalley material with large valley polarization and are very likely to be synthesized by the surface replacement technology \cite{Lu2017}. 

\indent In this work, the stability, valley polarization and magnetic anisotropy of Janus 2$H$-Gd$XY$ monolayers are systematically studied by the DFT calculations. All Janus 2$H$-Gd$XY$ monolayers show FM semiconductor characters with high T$_{c}$ beyond 260 K. Interestingly, only the 2$H$-GdBrCl monolayer possess the PMA behavior, while the 2$H$-GdICl and 2$H$-GdIBr monolayers exhibit the in-plane magnetic anisotropy (IMA) character. The competition between Gd atom-$p$/$d$ orbitals and halogen atom-$p$ orbital can result in a transition of easy axis direction from the [001] to [100] plane for the Janus 2$H$-Gd$XY$ monolayers. A spontaneous and robust valley polarization is observed, which can be effectively adjusted by the external biaxial strains. 

\section{Computational details}
All density functional theory (DFT) calculations are performed using the projected augmented wave (PAW) \cite{Kresse1999, E.1994} approach as implemented in the Vienna $ab$-$initio$ package (VASP) \cite{Kresse1996, J.1996}. Considering the exchange and correlation functional interactions, the Perdew-Burke-Ernzerhof (PBE) within generalized gradient approximation (GGA) is applied \cite{Perdew1996}. The van der Waals (vdW) correction is considered for the bulk Gd$XY$ using the Grimme (DFT-D3) method \cite{Grimme2010}. The plane wave cut-off energy is set to 500 eV and a vacuum space of 18 Å is applied along the z-axis [001] direction to avoid the interactions between adjacent layers. The crystal structure of the Janus 2$H$-Gd$XY$ monolayers is completely relaxed until a force of less than 0.01 eV/Å per atom and an energy difference of less than 10$^{-6}$ eV between two convergence steps is observed. The Brillouin zone is sampled using converged $\Gamma$-centered k-meshes with a density of 144 k-points (12$\times$12$\times$1) for structural relaxation and 576 k-points (24$\times$24$\times$1) for the electronic calculations \cite{Monkhorst1976}. The electron configurations including $5s^{2}5p^{6}4f^{7}5d^{1}6s^{2}$ for Gd \cite{You2021}, $4s^{2}4p^{5}$ for Br, $3s^{2}3p^{5}$ for Cl, and 5s$^{2}5p^{5}$ for I atom are considered. The SOC effect is included in the calculations to investigate electronic, magnetic and valley-related properties of the Janus 2$H$-Gd$XY$ monolayers. The rotationally invariant local spin density approximation (LSDA)+Hubbard ($U$) method is employed to treat the strongly correlated corrections to the Gd 4$f$ electrons \cite{Larson2007} and the corresponding on-site $U$/exchange interaction $J$ parameters is set at 9.20 eV/1.20 eV \cite{Wang2020, Jamnezhad2017}. The Phonon dispersion spectrum of the Janus 2$H$-Gd$XY$ monolayers are obtained by the PHONOPY code \cite{Togo2015, Gonze1997} using a 2$\times$2$\times$1 supercell. Ab initio molecular dynamic (AIMD) simulations \cite{Bucher2011} adopt the NVT ensemble \cite{Nos1984} based on the Nosé-Hothermostat \cite{Shuichi1984} controlled the temperature of systems at 300 K with a total of 8.0 $ps$ at 2.0 $fs$ per time step. The VASPKIT code is used to process some of the VASP data \cite{Wang2021}. The T$_{c}$ of the Janus 2$H$-Gd$XY$ monolayers are estimated by using the Monte Carlo simulation package MCSOLVER \cite{Liu2019} based on the Wolff algorithm. The Berry curvature and optical properties are calculated based on Fukui's method \cite{Fukui2005} by VASPBERRY code which is developed by Prof. Kim \cite{Kim2022}. 
\section{Results and discussion}
\begin{figure*}[hbt]
  \includegraphics[width=0.8\linewidth]{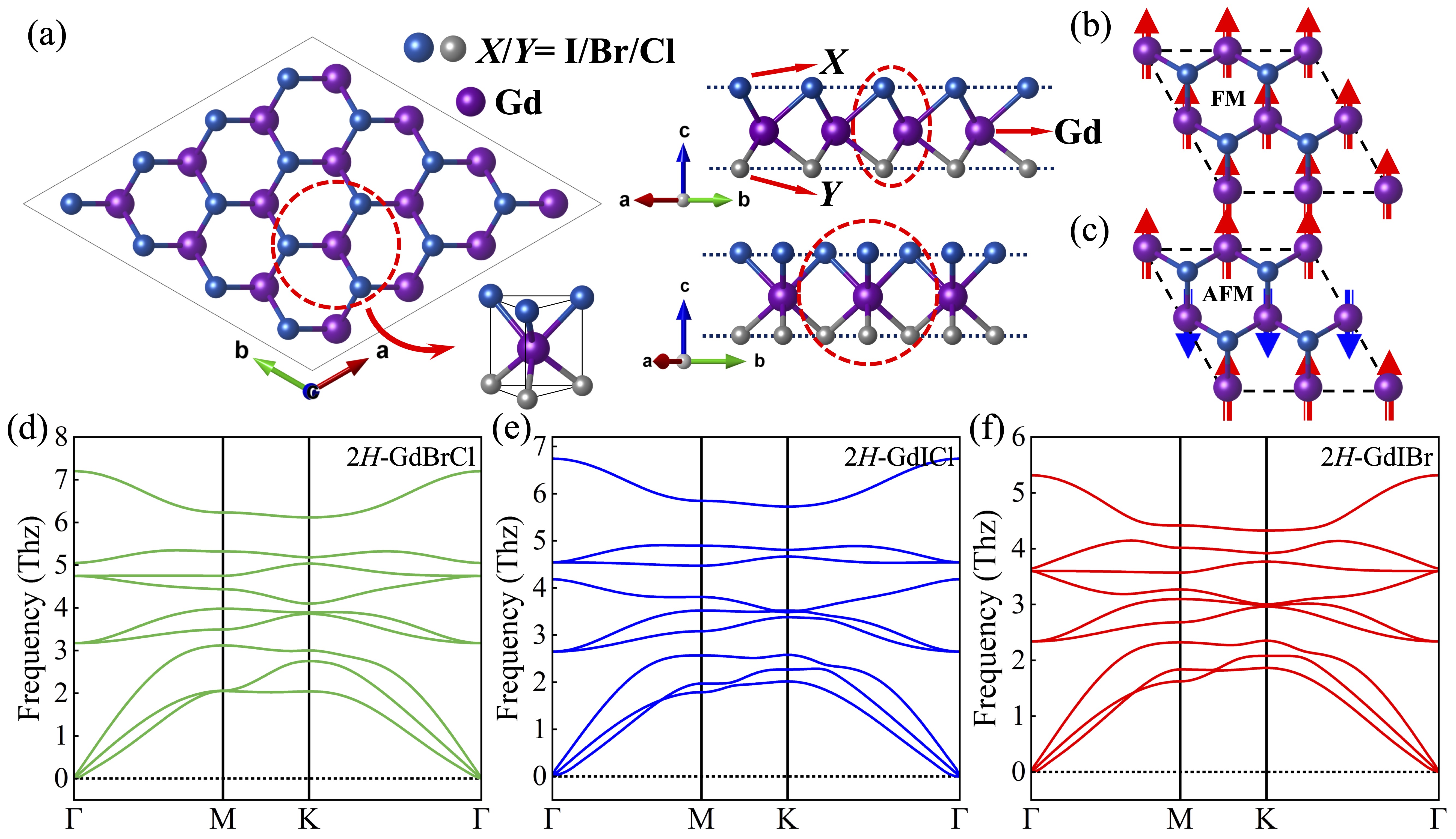} 
  \caption{ (a) Top/side views of the Janus 2$H$-Gd$XY$ monolayers and the coordination environment of the Gd atom. The unit cell is marked by the red lines. The (b) FM and (c) AFM states of crystal structures, red (blue) arrow shows the direction of spin-up (spin-down). The phonon dispersion of the Janus (d) 2$H$-GdBrCl, (e) 2$H$-GdICl, and (f) 2$H$-GdIBr monolayers.}\label{fig1}
\end{figure*}
\begin{figure*}[hbt]
  \centering
  \includegraphics[width=0.8\linewidth]{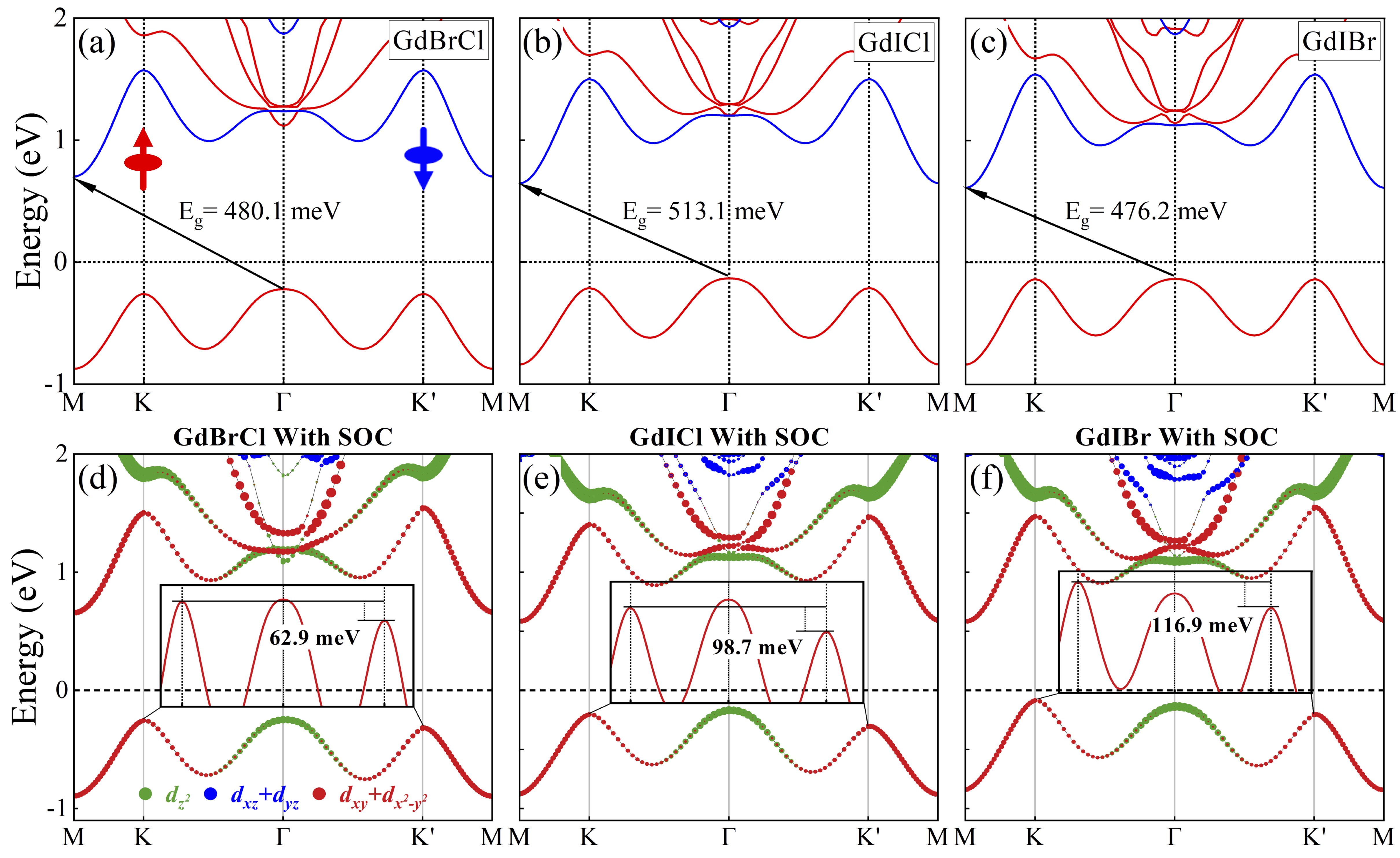} 
  \caption{The projected PBE band structures for the Janus 2$H$-Gd$XY$ monolayers (a-c) without and (d-f) with considering SOC. The Fermi level is set to zero.}\label{fig2}
\end{figure*}
\indent Figure\;\ref{fig1}(a) displays the crystal structure of the Janus 2$H$-Gd$XY$ monolayers. Clearly, the loss of reflection symmetry of Gd atom reduces the symmetry of the systems. The optimized lattice constants for the 2$H$-GdBrCl, 2$H$-GdICl, and 2$H$-GdIBr monolayers are 3.835, 3.96, and 4.019 Å, respectively. The FM and anti-ferromagnetic (AFM) ordering of crystal structures for the Janus 2$H$-Gd$XY$ monolayers is shown in Fig.\;\ref{fig1}(b) and \ref{fig1}(c). The ferromagnetic stability energies ($\Delta$E=E$_\mathrm{AFM}$$-$E$_\mathrm{FM}$) between the FM and AFM ordering for the 2$H$-GdBrCl, 2$H$-GdICl, and 2$H$-GdIBr monolayers are 165.2, 154.9, and 152.2 meV, respectively, strongly suggesting the existence of FM coupling among three systems. Figure\;S1 of the Supplemental Material (SM) \cite{SM2022} shows the spin density images of the Janus 2$H$-Gd$XY$ monolayers. One can see that the magnetic moments are mainly contributed by the Gd and Cl (Br) atoms for the Janus 2$H$-GdICl (2$H$-GdIBr) monolayers and by all Gd, Cl and Br atoms for the Janus 2$H$-GdBrCl monolayer, suggesting that the valence electrons tend to gather around the Cl (Br) atom with stronger electronegativity. In order to prove the stability of Janus 2$H$-Gd$XY$ monolayers, the calculations of phonon dispersion and AIMD simulation are carried out. As shown in Fig.\;\ref{fig1}(d)-\ref{fig1}(f), the phonon dispersions exhibit the positive value in the whole Brillouin zone, strongly suggesting the dynamical stability for the Janus 2$H$-Gd$XY$ monolayers. In addition, with time evolution, the small fluctuations (about $\pm$1 eV) of free energy, the total magnetic moment is kept at about 128.0 $\mu_{B}$ and the original configuration does not show large distortion have implied a good thermal stability of the Janus 2$H$-Gd$XY$ monolayers [Fig.\;S2(a)-S2(c) of the SM \cite{SM2022}]. To estimate the possibility of mechanical exfoliation, the exfoliation energy of the Janus 2$H$-Gd$XY$ monolayers is calculated in four-layer slab models with AB-stacking \cite{Li2021} [Fig.\;S3(a) of the SM \cite{SM2022}] from their layered bulk crystals. Considering the small value of separation distance, the exfoliation process is performed with the fixed atomic positions. As shown in Fig.\;S3(b)-S3(d) of the SM \cite{SM2022}, the increase of separation distance ($d-d_{0}$) leads to an obvious increase in the energy differences $\Delta$E=E$_{d}$$-$E$_{d_{0}}$, which converge to 0.232, 0.243, and 0.239 J/m$^{2}$, respectively, for the Janus 2$H$-GdBrCl, 2$H$-GdICl, and 2$H$-GdIBr monolayers, respectively. These cleavage energies are remarkably lower than the measured value (0.36 J/m$^{2}$) for the graphite \cite{Zacharia2004}, which can be further confirmed by the variation of cleavage strength (the first-order derivative of cleavage energy). This means that the Janus 2$H$-Gd$XY$ monolayers are easily to exfoliate experimentally.
\begin{figure}[hbt]
  \includegraphics[width=1.0\linewidth]{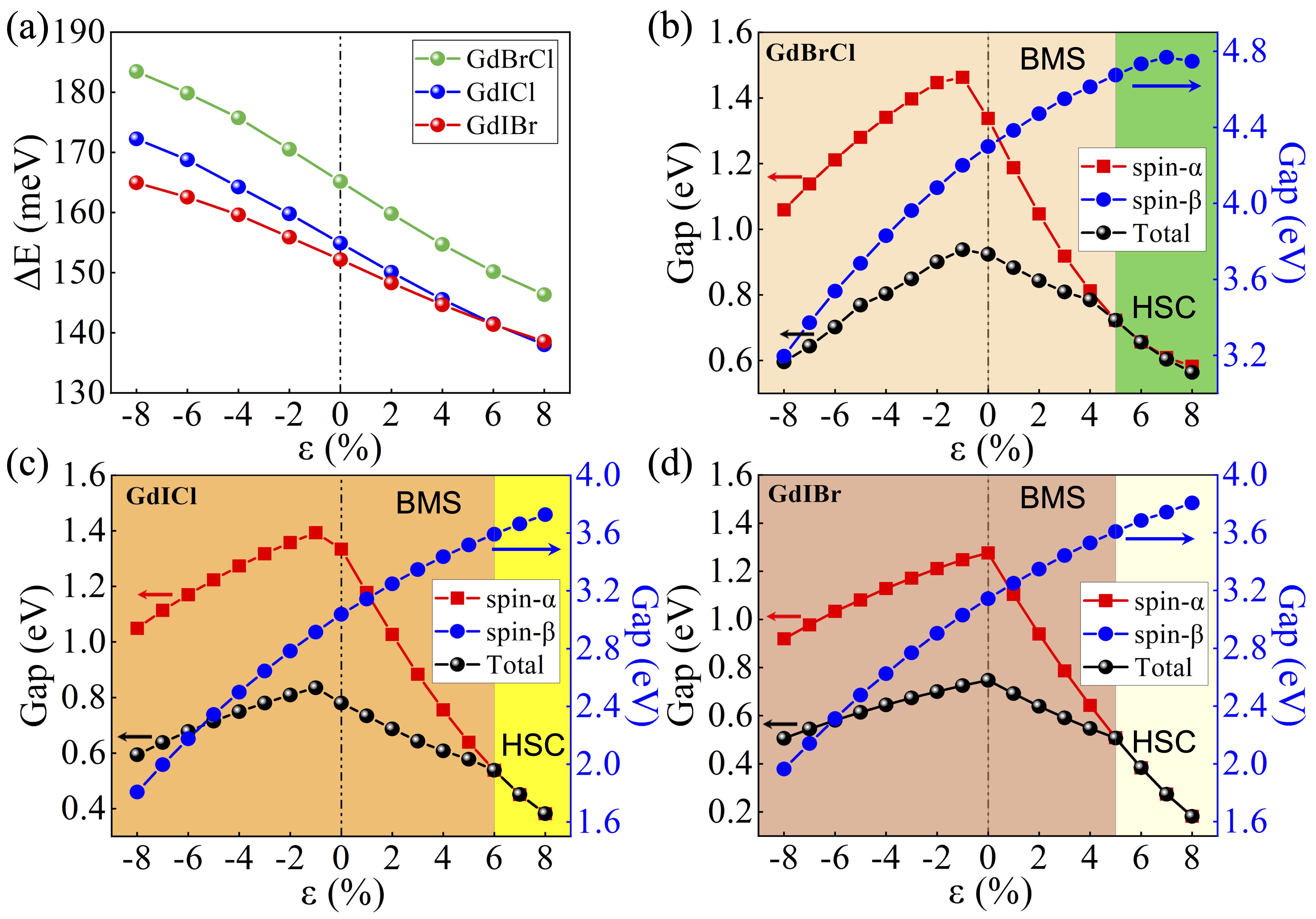} 
  \caption{(a) The biaxial strain dependences on the ferromagnetic stability energy $\Delta$E for the Janus 2$H$-Gd$XY$ monolayers. The biaxial strain dependences on E$_\mathrm{g}$-$\alpha$, E$_\mathrm{g}$-$\beta$, and E$_\mathrm{g}$ for the (b) 2$H$-GdBrCl, (c) 2$H$-GdICl, and (d) 2$H$-GdIBr monolayers.}\label{fig3}
\end{figure}
\begin{figure}[hbt]
  \includegraphics[width=1.0\linewidth]{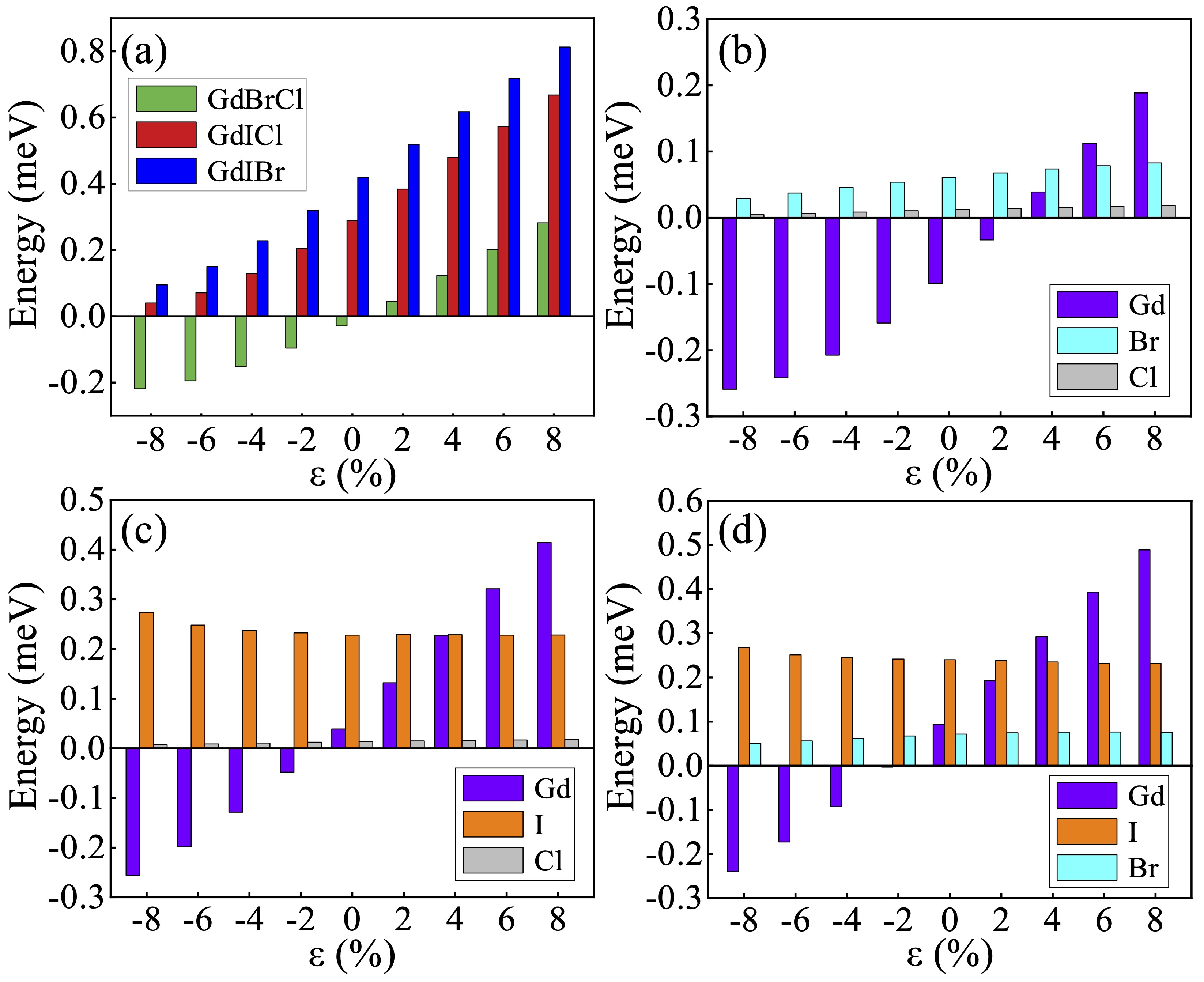} 
  \caption{The (a) total MAE and (b-d) atoms-resolved MAE for the Janus 2$H$-Gd$XY$ monolayers.}\label{fig4}
\end{figure}
\begin{figure*}[hbt]
  \centering
  \includegraphics[width=0.9\linewidth]{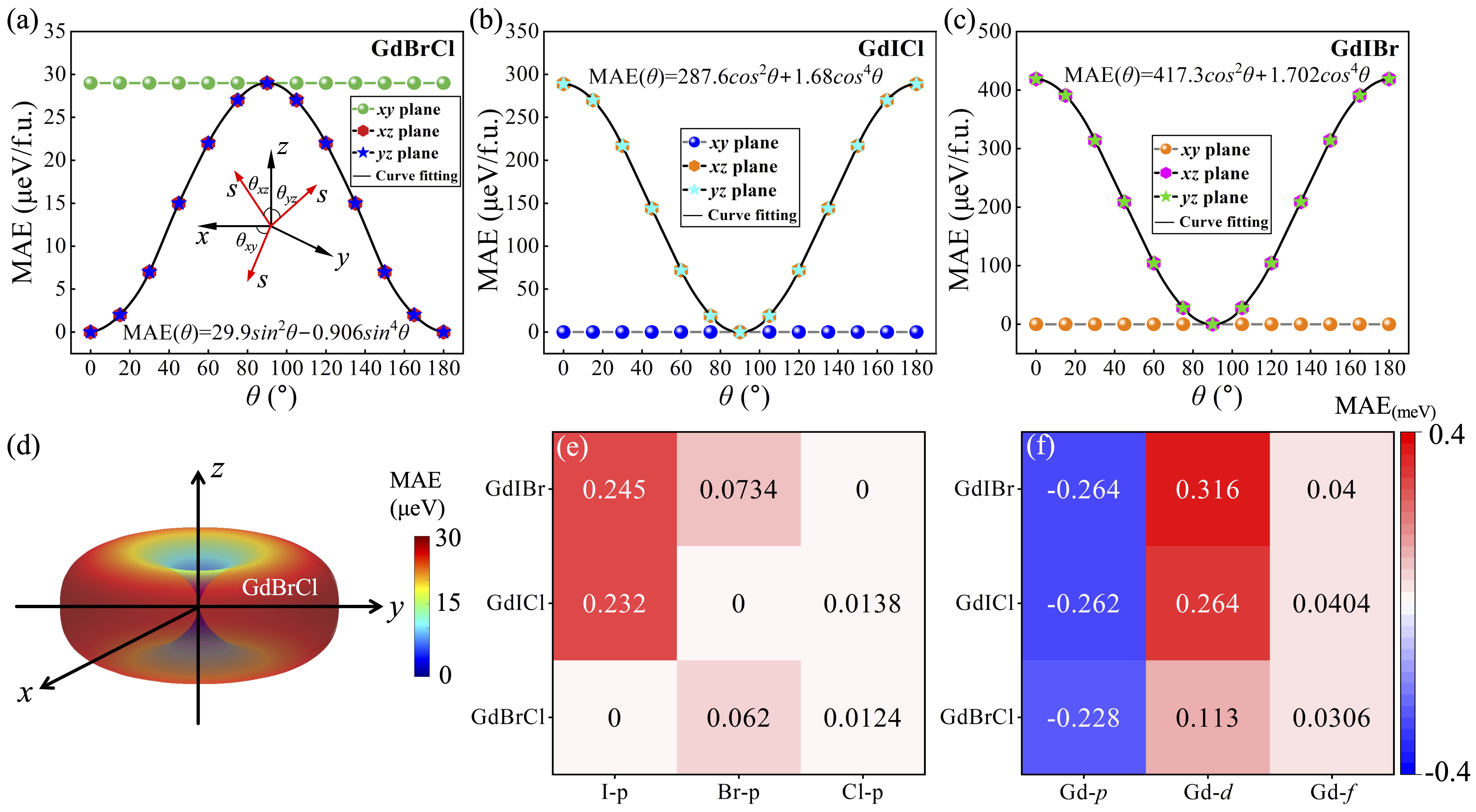} 
  \caption{MAE of the (a-c) Janus 2$H$-Gd$XY$ monolayers along $xy$, $xz$ and $yz$ planes and (d) over the whole space for the 2$H$-GdBrCl monolayer. The contributions of (e) Br/I/Cl-$p$ orbitals and (e) Gd-$p$, -$d$, -$f$ orbitals to MAE for the Janus 2$H$-Gd$XY$ monolayers.}\label{fig5}
\end{figure*}

The band structures of the Janus 2$H$-Gd$XY$ monolayers without considering SOC are shown in in Fig.\;\ref{fig2}(a)-\ref{fig2}(c). The energies at the K and K$^{\prime}$ valleys are equal, suggesting that the valley splitting does not appear without considering SOC. The typical bipolar magnetic semiconductor character [BMS: the valence and conduction bands possess opposite spin polarization when approaching the Fermi level (E$_\mathrm{F}$)] \cite{JG2012} with an indirect band gap can be observed, where the valence band maximum (VBM) located at the $\Gamma$ point and the conductor band minimum (CBM) located at the M point have the opposite spin polarization orientation. The band gaps for the Janus 2$H$-GdBrCl, 2$H$-GdICl, and 2$H$-GdIBr monolayers are about 480.1, 513.1, and 476.2 meV, respectively. Figure\;S4(a)-S4(c) of the SM \cite{SM2022} show the corresponding orbital-resolved density of states (DOS) curves of three systems. One can see that the VBM and CBM of the Janus 2$H$-Gd$XY$ monolayers are dominated by $d$ orbitals of Gd atoms. The narrow and high distribution of DOS for Gd-$f$ orbitals are around 3 eV above the E$_\mathrm{F}$. As a result, the 5$d$ electrons are fully spin-polarized in the vicinity of the E$_\mathrm{F}$ and the induced moments of 4$f$ electrons can effectively align around the polarized 5$d$ electrons by a cooperative mechanism, resulting in intrinsic large magnetic moments \cite{Roy2010}. On the contrary, the DOS of halogen atoms comes from their $p$-orbitals, and is far away from the E$_\mathrm{F}$, suggesting a weak hybridization between the $p$-orbitals of halogen atoms and 4$f$-orbitals of Gd atoms. Generally, it can consider that two magnetic exchange interaction mechanisms, namely the Gd-Gd direct-exchange interaction and the Gd-$X$($Y$)-Gd super-exchange interaction ($X$,$Y$ represent the halogen atoms), are responsible for the FM ordering in the Janus 2$H$-Gd$XY$ monolayers. Especially that the super-exchange FM (AFM) interaction occurs when the bond angle between cation and anion ions is close to 90$^{\circ}$ (180$^{\circ}$) by the Goodenough-Kanamori-Anderson rule \cite{Goodenough1955, Kanamori1960, Anderson1959}. It is noted that the bond angles of Gd-$X$($Y$)-Gd in the three systems almost reach to 90$^{\circ}$, namely Cl:86.645$^{\circ}$ (Br:82.018$^{\circ}$) in the 2$H$-GdBrCl, I:78.962$^{\circ}$ (Cl:89.159$^{\circ}$) in the 2$H$-GdICl, and I:80.18$^{\circ}$ (Br:85.70$^{\circ}$) in the 2$H$-GdIBr, strongly suggesting the existence of super-exchange FM interaction between the Gd 5$d$ and $X$($Y$)-4$p$ orbitals. In addition, the nearest Gd-Gd distance in the Janus 2$H$-Gd$XY$ monolayers is larger than 3.8 Å, which is much larger than the radii of $d$- or $f$- orbital carrying magnetic moments. Thus, based on above analysis, the super-exchange interaction is the dominant mechanism for the FM coupling between the nearest neighboring Gd cations mediated by $X$/$Y$ anions in Janus 2$H$-Gd$XY$ monolayers, which comes from the $p$-$d$ hybridization between the $X$/$Y$-4$p$ and Gd-5$d$ orbitals.

Figure\;\ref{fig2}(d)-\ref{fig2}(f) display the projected band structures of the Janus 2$H$-Gd$XY$ monolayers with considering SOC. Clearly, the K (K$^{\prime}$) valley still exist (same as the spin polarization), but the energies at the K and K$^{\prime}$ valleys are not equal and the degeneracy of the valley is broken. The energy at the K valley is higher than that at the K$^{\prime}$ valley in the VBM, resulting a large valley polarization of 62.9, 98.7, and 116.9 meV for the 2$H$-GdBrCl, 2$H$-GdICl, and 2$H$-GdIBr monolayers, respectively. Also, the K and K$^{\prime}$ valleys in the valence band mainly come from in-plane $d_{xy}$/$d_{x^{2}-y^{2}}$ orbitals and $\Gamma$ point is contributed by the out-of-plane $d_{z^{2}}$ orbital. Based on the above analysis, the electronic structures, valley and magnetic properties are expected to strongly depend on the occupation of in-plane atomic orbitals for the Janus 2$H$-Gd$XY$ monolayers.

Figure\;\ref{fig3}(a) shows the biaxial strain dependences on the ferromagnetic stability energy $\Delta$E for the Janus 2$H$-Gd$XY$ monolayers. Clearly, the ferromagnetic stability energy $\Delta$E gradually decreases with increasing the strain ($\varepsilon$) from $-$8\% to 8\%, suggesting that applying in-plane biaxial strain can well adjust the FM exchange interaction of studies systems. One can see from Fig.\;\ref{fig3}(b)-\ref{fig3}(d), and Fig.\;S5 of the SM \cite{SM2022} that with the increase of tensile strain, the band edge difference of CMB at the M and $\Gamma$ points between the spin-$\alpha$ (spin-up) and spin-$\beta$ (spin-down) electrons can be expressed as $\Delta$E$_\mathrm{CBM-\Gamma-M}$ = E$_\mathrm{CBM-\Gamma-\alpha}$$-$E$_\mathrm{CBM-M-\beta}$. For the unstrained Janus 2$H$-GdBrCl, 2$H$-GdICl and 2$H$-GdIBr monolayers, $\Delta$E$_{CBM-\Gamma-K}$ = 0.4135, 0.5545 and 0.5286 eV, respectively. Clearly, the E$_\mathrm{g}$-$\alpha$ and E$_\mathrm{g}$ show a monotonous decrease trend, while the E$_\mathrm{g}$-$\beta$ increases monotonously. When the tensive strain (about 5-6\%) is applied, the $\Delta$E$_\mathrm{CBM-\Gamma-K}$ is up to 0 eV, indicating a transition from BMS to half-semiconductors (HSC) \cite{Guan2020} characters for the Janus 2$H$-Gd$XY$ monolayers, namely, both VBM and CBM are occupied by the same spin direction elections (spin-$\alpha$). As the tensile strain increases continuously, the Janus 2$H$-Gd$XY$ monolayers still keep HSC character. Interestingly, this transition from BMS to HSC characters does not appear under compressive strain due to the occupation of spin-$\alpha$ electrons. Under the biaxial strains varying from $-$8\% to 0\%, the Janus 2$H$-Gd$XY$ monolayers still maintain the BMS character due to the occupation of spin-$\beta$ electrons in the CBM, which shows a monotonous decrease with strain for the spin-down band gap E$_\mathrm{g}$-$\beta$. 

Magnetic anisotropy (MA) is one key factor to realize the long-range magnetic ordering in 2D materials, which directly correlates with the thermal stability of magnetic data storage. The MA can be scaled by magnetic anisotropy energy (MAE), which is expresses as MAE = E$_\mathrm{[001]}$$-$E$_\mathrm{[100]}$,  where E$_\mathrm{[001]}$ and E$_\mathrm{[100]}$ are the energies of system when the magnetic moments are along with the out-of-plane [001] axis and the in-plane [100] axis, respectively. The MAE values for the Janus 2$H$-GdICl and 2$H$-GdIBr monolayers are 0.289 and 0.419 meV, respectively, showing the in-plane magnetic anisotropy (IMA) character. Moreover, the MAE value of $-$0.029 meV implies a perpendicular magnetic anisotropy (PMA) character for the Janus 2$H$-GdBrCl monolayer. Figure\;\ref{fig4}(a) shows the biaxial strain dependences on total MAE of Janus 2$H$-Gd$XY$ monolayers, which exhibits a monotonous increase behavior with increasing the strain from $-$8\% to 8\%. The biaxial strain dependences on atomic-resolved MAEs are shown in Fig.\;\ref{fig4}(b)-\ref{fig4}(d). The halogen $X$ ($Y$) atom in the Janus 2$H$-Gd$XY$ monolayers always keeps the IMA behavior at the whole strain ranges. Moreover, the MAE from the Gd atom increases monotonously with the increase of tensile or compressive strains, and a transition from the PMA to IMA behavior can be observed. Figure\;\ref{fig5}(a)-\ref{fig5}(c) show the MAE of the Janus 2$H$-Gd$XY$ monolayers along $xy$, $xz$ and $yz$ planes. Clearly, for all Janus 2$H$-Gd$XY$ monolayers, the MAE exhibits a strong dependence of magnetization direction along the $xz$ and $yz$ planes. Further, the MAE as a function of angle $\theta$ can be expressed as \cite{Ethem2021}: 
\begin{align}
  \begin{split}
      \operatorname{MAE}(\theta)=K_1 \sin ^2 \theta+K_2 \sin ^4 \theta
  \end{split}
\end{align}
and
\begin{align}
  \begin{split}
      \operatorname{MAE}(\theta)=K_3 \cos ^2 \theta+K_4 \cos ^4 \theta
  \end{split}
\end{align}
where $K_{1}$ ($K_{2}$) and $K_{3}$ ($K_{4}$) are the anisotropy constants. The angle theta $\theta$ refers to an arbitrary azimuthal angle $\theta$ [0$^{\circ}$, 180$^{\circ}$]. It is clear that the MAE is independent of the angle $\theta_{xy}$ ($xy$ plane). For the 2$H$-GdBrCl monolayer, the best fitting data ($K_{1}$>0 and $K_{1}$$\gg$$K_{2}$) of MAE dependence on $\theta$ show a magnetization direction along out-of-plane axis. While, for the 2$H$-GdICl and 2$H$-GdIBr monolayer, both $K_{3}$ and $K_{4}$ are positive, revealing that the MAE of the two systems prefer a single easy axis. The strong magnetic anisotropy of the three systems is also confirmed by the distribution of MAE on the whole spaces [Fig.\;\ref{fig5}(d) and Fig.\;S6 of the SM \cite{SM2022}]. The calculated orbital-resolved MAE based on the spin-orbit matrix element differences \cite{Wang1996} is used to investigate its origins and can be expressed as: 
\begin{align}
  \begin{split}
      \mathrm{MAE}=\xi^2 \sum_{o, u} \frac{\left|\left\langle\psi_o\left|\hat{L}_z\right| \psi_u\right\rangle\right|^2-\left|\left\langle\psi_o\left|\hat{L}_x\right| \psi_u\right\rangle\right|^2}{E_u-E_o}
  \end{split}
\end{align}
where $E_{o}$ ($E_{u}$) represents the energy of the occupied (non-occupied) state and the $\xi$ is the SOC constants. As is shown in Fig.\;\ref{fig5}(e) and \ref{fig5}(f), for the Janus 2$H$-Gd$XY$ monolayers, the negative MAE is contributed by the Gd-$p$ orbital. On the contrary, the sizable positive MAE comes from the contribution of Gd-$d$ and I-$p$ orbitals. Other Gd-$f$ and Br/Cl-$p$ orbitals only give an almost negligible contributions to the MAE. Therefore, the difference in the total MAE for the Janus 2$H$-Gd$XY$ monolayers is a competitive result of the Gd-$d$/-$p$ and $X$ ($Y$)-$p$ orbitals. As the $X$ ($Y$) changes from the I to Br (Cl) atom, the net value of total MAE of $p$ orbitals of $X$ ($Y$) atoms decrease, resulting in out-of-plane MA for the Janus 2$H$-GdBrCl monolayer. \\
\indent T$_{c}$ is an important indicator for the potential applications of magnetic devices. In order to evaluate the T$_{c}$ of the Janus 2$H$-Gd$XY$ monolayers, the Metropolis Monte Carlo simulations are carried out based on the Heisenberg model. Following the previous work \cite{Liu2021}, we only considered the nearest Gd-Gd magnetic interaction. The spin Hamiltonian can be described with the following formulas:
\begin{align}
  \begin{split}
      H=-J \sum_{i, j} \vec{S}_i \cdot \vec{S}_j-D\left(S_i^z\right)^2
  \end{split}
\end{align}
where $J$ represents the magnetic exchange parameter, $S$ is the spin vector, and $D$ is an anisotropy energy parameter.
\begin{align}
  \begin{split}
      \mathrm{E}_{\mathrm{FM}}=\mathrm{E}_0-6 J|S|^2-D|S|^2 
  \end{split}
\end{align}
\begin{align}
  \begin{split}
\mathrm{E}_{\mathrm{AFM}}=\mathrm{E}_0+2 J|S|^2-D|S|^2
  \end{split}
\end{align}
where the E$_\mathrm{FM}$ and E$_\mathrm{AFM}$ represent the total energy of systems with FM and AFM ordering. The magnetic exchange parameter $J$ with considering nearest neighbor coordination can be expressed using the following formula:
\begin{align}
  \begin{split}
      J=\frac{\mathrm{E}_{\mathrm{AFM}}-\mathrm{E}_{\mathrm{FM}}}{8|S|^2}
  \end{split}
\end{align}
\begin{figure}[hbt]
  \includegraphics[width=1.0\linewidth]{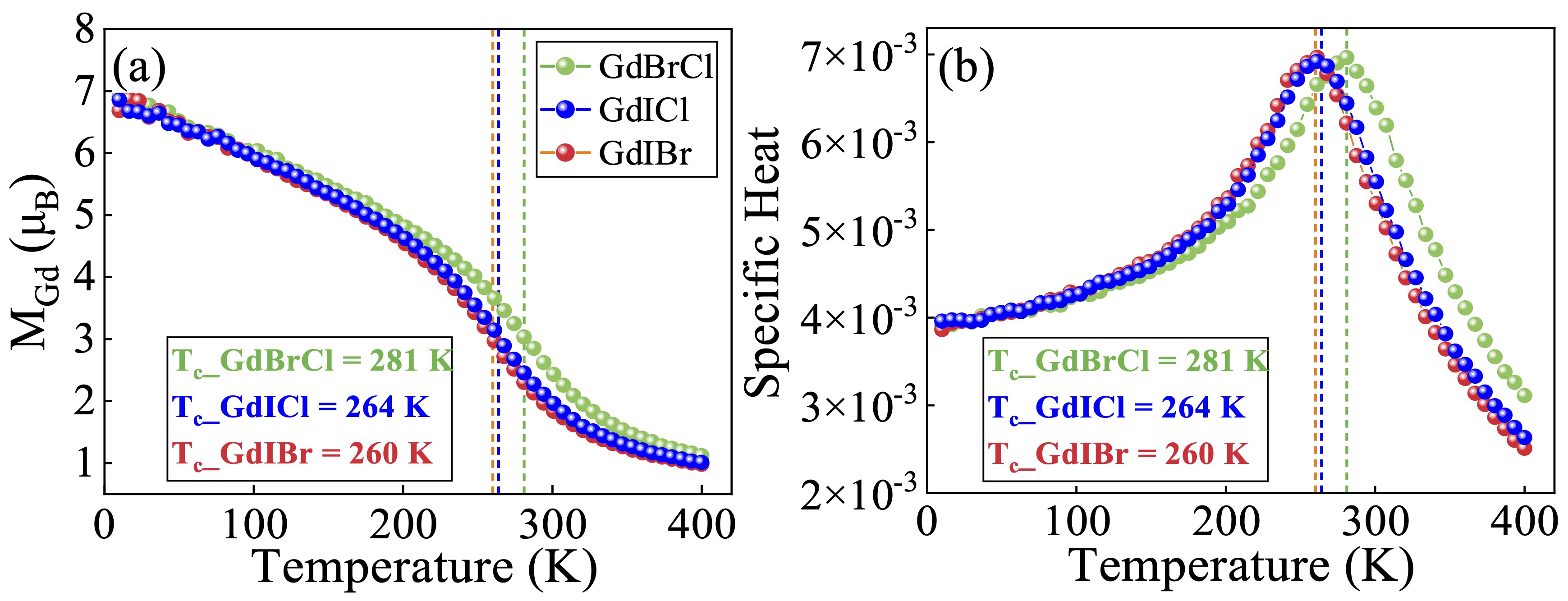} 
  \caption{(a) The magnetic moments (M$_\mathrm{Gd}$) and (b) specific heat for the Janus 2$H$-Gd$XY$ monolayers as functions of temperature.}\label{fig6}
\end{figure}
\begin{figure}[hbt]
  \includegraphics[width=1.0\linewidth]{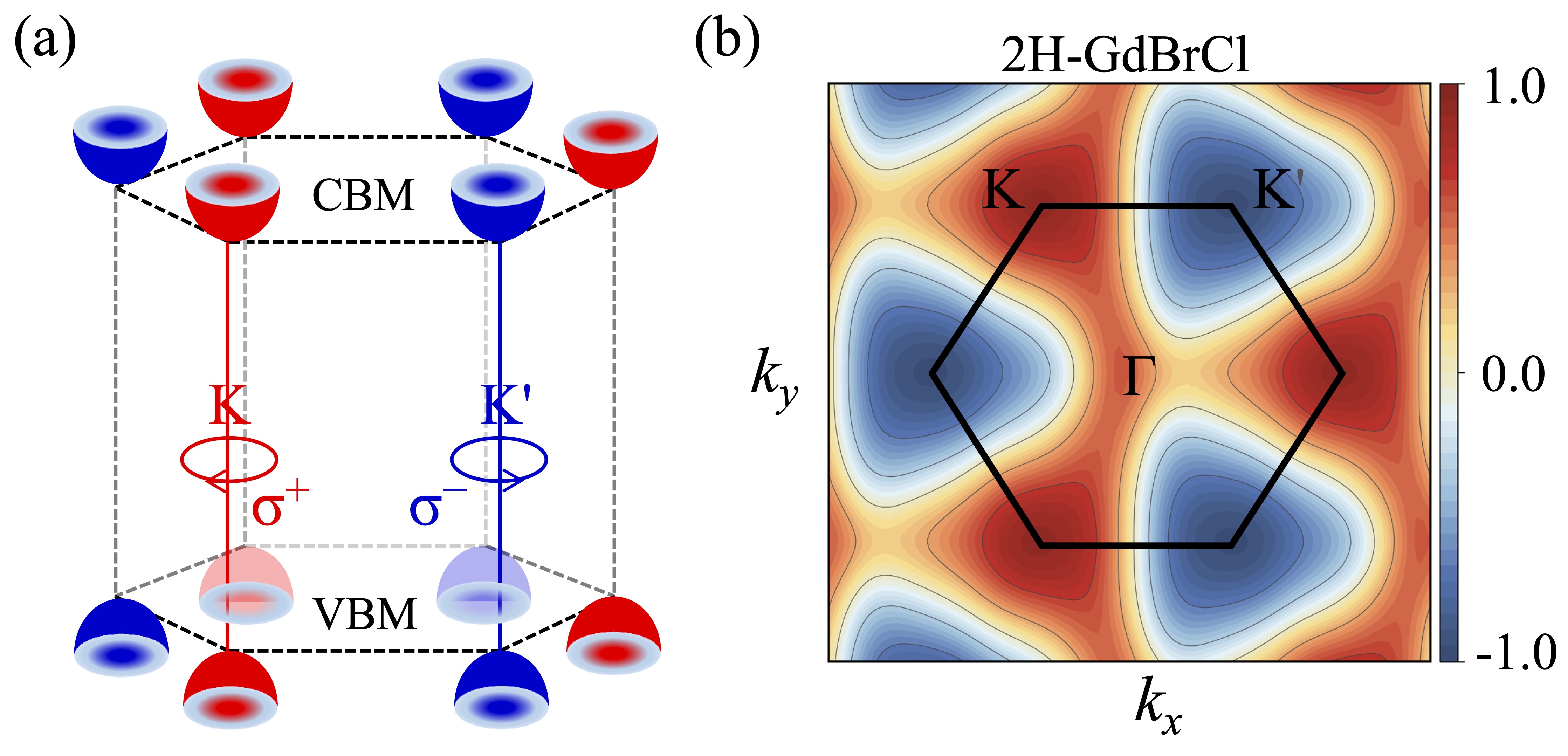} 
  \caption{(a) The valley optical transitions with $\sigma^{+}$ ($\sigma^{-}$) circular polarizations at the K (K$^{\prime}$) valleys. (b) The circular polarization $\eta(k)$ of optical transition for the Janus 2$H$-GdBrCl monolayer over the first Brillouin zone.}\label{fig7}
\end{figure}
\begin{figure*}[hbt]
  \centering
  \includegraphics[width=0.8\linewidth]{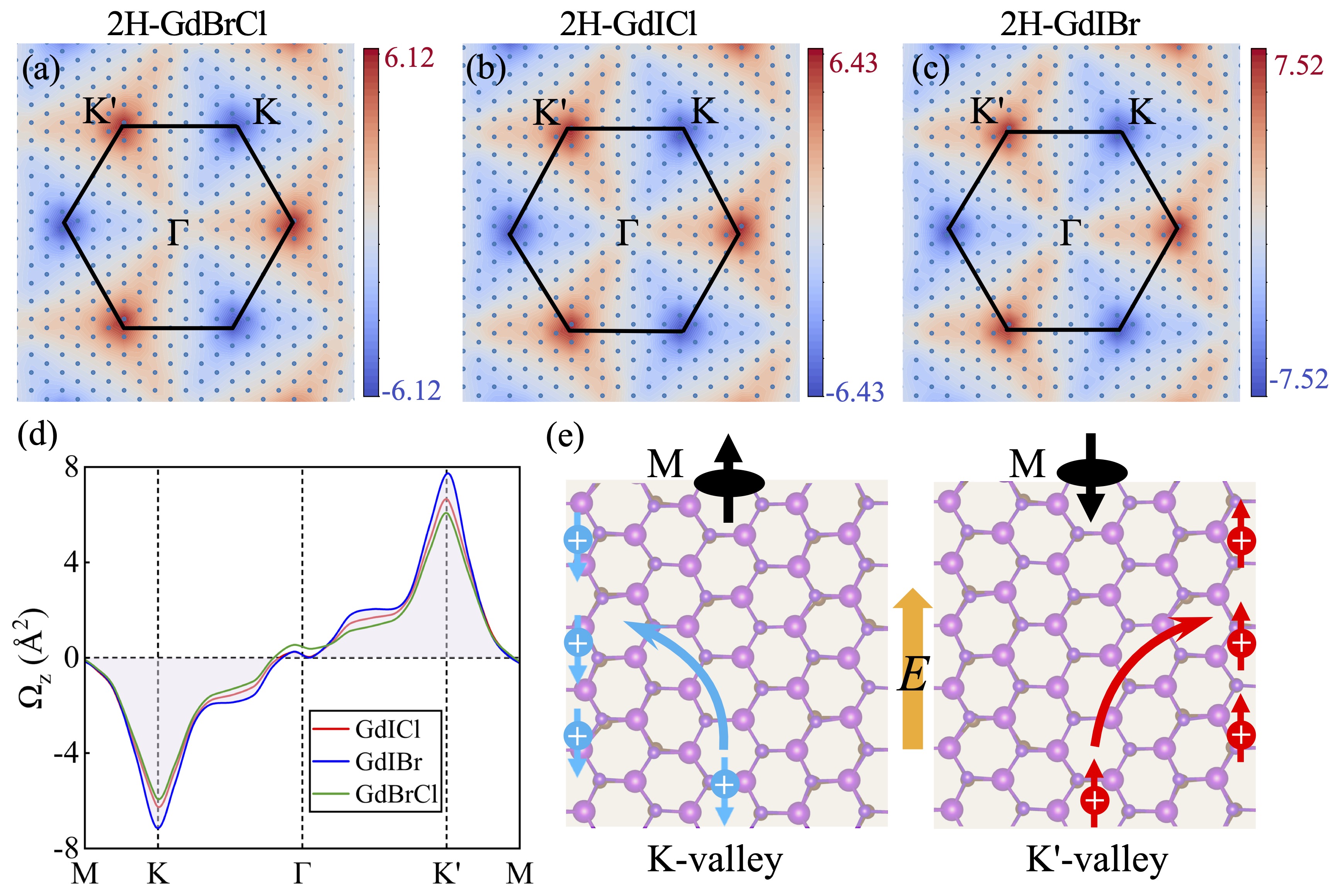} 
  \caption{(a-c) Contour map of Berry curvatures as well as (d) Berry curvatures along the high symmetry points for the Janus 2$H$-Gd$XY$ monolayers. (e) Schematic diagram of anomalous valley Hall effects under external electric fields $E$. The red and blue arrows represent the spin-up and spin-down holes.}\label{fig8}
\end{figure*}
The calculated $J$ values for the Janus 2$H$-GdBrCl, 2$H$-GdICl, and 2$H$-GdIBr monolayers are 1.29, 1.21, and 1.188 meV, respectively, strongly suggesting the existence of FM ordering around the nearest neighbor Gd atom. Figure\;\ref{fig6}(a) and \ref{fig6}(b) show the magnetic moments (M$_\mathrm{Gd}$) and specific heat for the Janus 2$H$-Gd$XY$ monolayers as functions of temperature. The calculated T$_{c}$ values are about 281, 264, and 260 K, respectively, for the Janus 2$H$-GdBrCl, 2$H$-GdICl, and 2$H$-GdIBr monolayers, which are obviously higher than those observed in the CrI$_{3}$ (45 K) \cite{Huang2017} and VI$_{3}$ (49 K) \cite{Kong2019} monolayers. \\
\indent Figure\;\ref{fig7}(a) shows the valley optical transitions with $\sigma^{+}$ ($\sigma^{-}$) circular polarizations at the K (K$^\prime$) valleys, which describes the intrinsic transport arising from the electronic structures of unequal K(K$^{\prime}$) valley. Further, the carriers of K and K$^{\prime}$ valleys can be selectively excited by the valley-selective circular dichroism with different chiral lights, which has been proved in the non-magnetic MoS$_{2}$ system \cite{Feng2012}. Inspired by this, for investigating corresponding optical properties, the degree of circular polarization is calculated by:
\begin{align}
  \begin{split}
      \eta(k)=\frac{\left|p_{+}^{V, C}(k)\right|^2-\left|p_{-}^{V, C}(k)\right|^2}{\left|p_{+}^{V, C}(k)\right|^2+\left|p_{-}^{V, C}(k)\right|^2}
  \end{split}
\end{align}
where $p_{\pm}^{V, C}(k)=p_x^{V, C}(k) \pm i p_y^{V, C}(k)$ and the matrix element is given by $p_{\pm}^{V, C}(k)=\left\langle\psi_k^V\left|\hat{p}_{\pm}\right| \psi_k^C\right\rangle$. The k-resolved quantity distinguishes left ($\sigma^{+}$) and right ($\sigma^{-}$) circularly polarized light between the valence and conduction bands with band index V and C, respectively. Figure\;\ref{fig7}(b) shows the circular polarization $\eta(k)$ of optical transition for the Janus 2$H$-GdBrCl monolayer over the first Brillouin zone. It can be found that $\eta(k)$ is nearly +1 ($-$1) at the K and K$^{\prime}$ valleys, indicating that left ($\sigma^{+}$) and right ($\sigma^{-}$) circularly polarized light can only excite the corresponding K and K$^{\prime}$ valley, respectively. Meanwhile, the Berry curvature of the matrix element $p_{\pm}^{V, C}(k)$ is given by \cite{Yao2008, Chang1996}:
\begin{align}
  \begin{split}
      \Omega_n(k)=-\frac{\hbar^2}{2 m} \sum_{V \neq C} \frac{\eta^{V, C}(k)\left[\left|p_{+}^{V, C}(k)\right|^2+\left|p_{-}^{V, C}(k)\right|^2\right]}{\left[E_C(k)-E_V(k)\right]^2}
  \end{split}
\end{align}
Clearly, the transformation property of $\eta^{V, C}(k)$ follows that of $\Omega_n(k)$, which can be confirmed by our following work.

After clarifying the spontaneous valley polarization in Janus 2$H$-Gd$XY$ monolayers, we finally consider their Berry curvatures along the out-of-plane direction [001] to describe their electronic transport properties, which is critical for quantum Hall effects. The Berry curvature of Janus 2$H$-Gd$XY$ monolayers can be defined as \cite{Xiao2010, Guido2018, Liu2020}: 
\begin{align}
  \begin{split}
  \Omega_n(k)=\nabla \times C_n(k)
  \end{split}
\end{align}
\begin{align}
  \begin{split}
  C_n(k)=i\left\langle u_n(k)\left|\nabla_k\right| u_n(k)\right\rangle
  \end{split}
\end{align}
where $n$ is the band index and $u_n(k)$ is the Bloch wave functions. For the Janus 2$H$-Gd$XY$ monolayers, all valence electrons below the E$_\mathrm{F}$ are considered during the calculation of the Berry curvature. The Berry curvatures of the Janus 2$H$-Gd$XY$ monolayers over the whole 2D Brillouin zone and along the high-symmetry lines are shown in Fig.\;\ref{fig8}(a)-\ref{fig8}(d). It is clear that the intensity of Berry curvatures at the K$^{\prime}$ valley is larger than that at the K valley, in accompanied with the opposite signs. Interestingly, two valleys in valence band are not only distinguishable in terms of energy, but also can be distinguished by the value of their Berry curvature. Figure\;\ref{fig8}(e) shows the schematic diagram of anomalous valley Hall eﬀects under external electric fields $E$. One can see that when the Janus 2$H$-Gd$XY$ monolayers are magnetized along the z-direction, the spin-down holes from the K$^{\prime}$ valley move to the left-hand side of the sample and are further accumulated on the boundary under the external electric field $E$. Conversely, when the magnetic field direction is altered along the $-$z-direction, the spin-up holes from the K valley move to the right-hand side of the sample and are accumulated on another boundary due to the opposite Berry curvature under the same electric field $E$. Therefore, the valley degree of freedom of carriers in the Janus 2$H$-Gd$XY$ monolayers can be selectively manipulated by the electric measurements, which is significance for designing the valleytronic devices. An external biaxial strain is applied on the Janus 2$H$-Gd$XY$ monolayers to investigate the robustness of the ferro-valley characters. Fig.\;S7(a) and S7(b) of the SM \cite{SM2022} show the variation of valley splitting of the Janus 2$H$-Gd$XY$ monolayers as a function of biaxial strains from $-$8\% to 8\%. Obviously, as the tensile (compressive) strains increase from 0\% to 8\% ($-$8\%), the valley splitting shows a monotonical increases (decreases) behavior. The valley polarization values of Janus 2$H$-GdBrCl, 2$H$-GdICl and 2$H$-GdIBr monolayer reaches 113.1, 142.7 and 158.8 meV at tensile strain of 8\%, respectively. Conversely, under a compressive strain of 8\%, valley polarization values decrease to 16.8 and 36.7 meV for the 2$H$-GdICl and 2$H$-GdIBr monolayers, respectively, implying a robust valley polarization against the biaxial strains. Interestingly, the energies of the K and K$^{\prime}$ valleys reverse for the 2$H$-GdBrCl monolayer, in accompanied with a valley splitting of $-$20.8 meV. Therefore, the Janus 2$H$-Gd$XY$ monolayers are excellent ferro-valley materials with large and tunable valley splitting, which is necessary to readily access and manipulate valleys for memory and logic applications. 
\section{Conclusions}
In summary, by first principles calculations, the electronic structure, valley polarization and MA of the Janus 2$H$-Gd$XY$ monolayers are systematical investigated, which exhibit FM semiconductor character with excellent dynamic/thermal stability. The estimated T$_{c}$ based on the Monte Carlo simulations are 284, 264, and 260 K for the 2$H$-GdBrCl, 2$H$-GdICl, and 2$H$-GdIBr monolayers, respectively. The magnetic easy axis direction of 2$H$-GdICl and 2$H$-GdIBr monolayers is along the in-plane direction. However, a transition from IMA to PMA occurs for the 2$H$-GdBrCl monolayer, which can be attributed due to the competition between the contributions of Gd-$p$/-$d$ and $X$($Y$)-$p$ orbitals to MAE. In addition, the valley polarization of Janus 2$H$-Gd$XY$ monolayers are robust against the external strains. Moreover, due to the breaking of time-reversal and inversion symmetry, nonzero out-of-plane Berry curvature is observed for the Janus 2$H$-Gd$XY$ monolayers, which makes it possible to realize the anomalous valley Hall effect. Overall, our work provides a new platform for the development of spin/valleytronic devices.

\begin{acknowledgments}
  This work was supported by the Natural Science Foundation of Tianjin City (Grant No. 20JCYBJC16540).
\end{acknowledgments}

\end{document}